\documentclass[twocolumn,english,superscriptaddress,floatfix]{revtex4}
\usepackage{graphicx}
\usepackage{bm, amsmath, amssymb}
\usepackage{color}
\usepackage{soul}
\usepackage{pdfpages}
\usepackage[T1]{fontenc}
\usepackage[latin9]{inputenc}
\usepackage{amstext}
\usepackage{bbold}
\usepackage{esint}
\usepackage{epsfig}
\usepackage{ellipsis}
\usepackage{lipsum}   
\usepackage{titlesec}  

\usepackage{fancyhdr}  

\usepackage{amssymb}
\usepackage{amsthm}
\usepackage{graphicx} 
\usepackage{epstopdf}
\usepackage{pdflscape} 
\usepackage{booktabs} 
\usepackage{array} 
\usepackage{paralist} 
\usepackage{verbatim} 
\usepackage{subfig} 
\usepackage{color}
\usepackage{caption}
\usepackage{fancyref }

\usepackage{newfloat}
\DeclareFloatingEnvironment[name={Supplementary Figure}]{suppfigure}

\usepackage{babel}

\usepackage{hyperref}
\hypersetup{
    colorlinks=true,
    linkcolor=blue,
    filecolor=magenta,      
    urlcolor=cyan,
    pdftitle={Sharelatex Example},
    bookmarks=false,
    pdfpagemode=FullScreen,
}
 
\usepackage[figurename=Fig.]{caption} 

\DeclareCaptionJustification{justified}{\justifying}
\usepackage{ragged2e}
\DeclareCaptionJustification{justified}{\justifying}
\begin{document}

\title{New hallmarks of criticality in recurrent neural networks}

\author{Yahya Karimipanah, Zhengyu Ma and Ralf Wessel \\ Department of Physics, Washington University, St.\ Louis, Missouri 63130 }


\begin{abstract}

A rigorous understanding of brain dynamics and function requires a conceptual bridge between multiple levels of organization, including neural spiking and network-level population activity. Mounting evidence suggests that neural networks of cerebral cortex operate at criticality. How operating near this network state impacts the variability of neuronal spiking is largely unknown. Here we show in a computational model that two prevalent features of cortical single-neuron activity, irregular spiking and the decline of response variability at stimulus onset, are both emergent properties of a recurrent network operating near criticality. Importantly, our work reveals that the relation between the irregularity of spiking and the number of input connections to a neuron, i.e., the in-degree, is maximized at criticality. Our findings establish criticality as a unifying principle for the variability of single-neuron spiking and the collective behavior of recurrent circuits in cerebral cortex.
\end{abstract}

\maketitle
\protect \section{Introduction}
Linking the evolutionary-derived multi-scale organization of the brain to neural dynamics and computation represents a major challenge in systems neuroscience \cite{sompolinsky2014computational}. Among the vast spectrum of spatial and temporal scales of brain activity, two experimentally accessible levels of brain organization are (i) the single-neuron spiking and (ii) the population activity of the network in which the neurons are embedded to various degrees. Single-neuron spiking in cerebral cortex is characterized by statistical properties, such as irregular spiking \cite{softky1993highly} and reduced variability during sensory stimulation \cite{churchland2010stimulus}. Population activity is characterized by complex spatiotemporal activity, including scale-free activity \cite{schuster2014criticality}, which is predicted to occur for a network state near criticality \cite{bak1987self}. These observations at two adjacent levels of brain organization raise the question, to what extent the network state controls the variability of single-neuron spiking?

Irregular spiking, defined as the mean {\it coefficient of variation} (CV) being larger than one, is known as one of the most widespread features of cortical activity in vivo \cite{softky1993highly,bellay2015irregular}. Theoretical studies have linked the prevalence of irregular spiking to the fluctuations of synaptic inputs at the sub-threshold regime \cite{deneve2016efficient,gerstner2014neuronal}. Despite the popularity of this hypothesis, here we provide an alternative scenario, by which the onset of irregular spiking can simply emerge at the transition between two phases of order and chaos. Further, using a computational model we show that even at the presence of other mechanisms for irregular spiking ($CV>1$), neuronal activity shows maximum irregularity when the network resides at criticality. Moreover, in addition to irregular spiking, we show that criticality also gives rise to pronounced decline in neural variability after the onset of stimulus, as well as maximum correlation between a number of single-neuron properties such as their CV's and firing rates. Our findings provide us with robust measures of critical dynamics that at the same time could further our understanding about the implications of criticality for brain dynamics and function. 

\section{Results} 
\begin{figure*}
\centering
\includegraphics[scale=0.85]{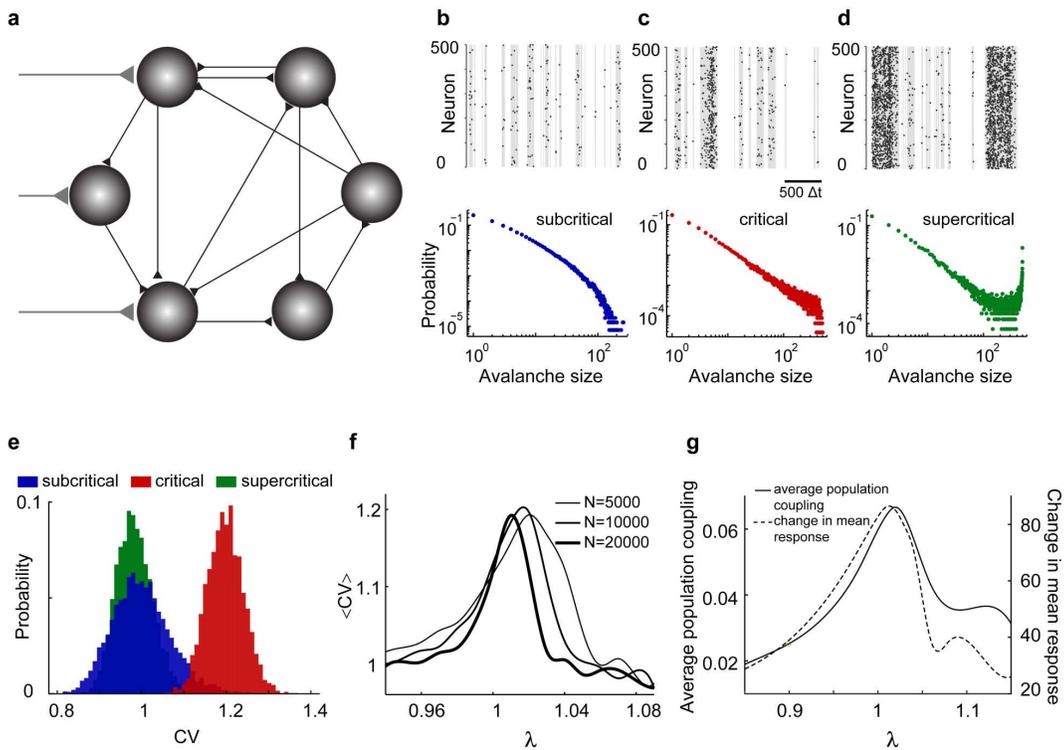}
\caption[{\protect \footnotesize Irregular spiking emerges in a recurrent network operating at the critical point:}]{\protect \footnotesize {{\bf Irregular spiking emerges in a recurrent network operating at the critical point:} } {\bf (a)}    The model network consists of binary probabilistic model neurons with sparse connectivity (black) and weak external inputs (gray) to a fraction of the neurons. {\bf (b-d)} Simulated spike trains (black raster), neuronal avalanches (gray), and corresponding avalanche size distributions (for $5 \times 10^5$ simulation time-steps) for a network of $N=500$ neurons with $10\%$ connectivity and external inputs $\eta=1/(10N)$ to all neurons. Simulations were conducted for three different network states: subcritical ((b) $\lambda=0.9$, blue), critical ((c) $\lambda=1.0$, red), and supercritical ((d) $\lambda=1.1$, green). {\bf (e)} Inter-spike-interval CV distributions of simulated spike trains from a network of $N=5000$ neurons with $3\%$ connectivity and external inputs $\eta=1/(5N)$ to all neurons. Simulations were conducted for the subcritical ($\lambda=0.9$, blue), critical ($\lambda=1.02$, red), and supercritical ($\lambda=1.06$, green) network states. {\bf (f)} The average CV as a function of the maximum eigenvalue $\lambda$ of the transition probability matrix $P_{ij}$ for three network sizes. Connectivity was $3\%$ and external input was $1/(5N)$ to all neurons. The curves were based on 13 values of $\lambda$ within the range shown and were smoothed using Matlab spline. {\bf (g)} The average population coupling and the change in mean response as a function of $\lambda$ for a network of $N=5000$ neurons with $3\%$ connectivity and external inputs of strength $\eta=1/(5N)$ applied to all neurons. To compute the change in mean response, we increased the external input strength by a factor of $10$ half-way through the simulation, i.e., from $\eta=1/(5N)$ to $\eta=2/N$. 
}
\label{fig1}
\end{figure*}

To explore the impact of critical dynamics on the statistics of single-neuron spiking, we used a model network consisting of excitatory binary probabilistic neurons with sparse connectivity and external inputs (see fig.\ref{fig1}a and \nameref{methods}). This model permits the use of the maximum eigenvalue $\lambda$ of the {\it transition probability matrix} $P_{ij}$ as a control parameter to tune the network near the critical point \cite{larremore2011predicting,larremore2012statistical}. Such tuning results in characteristic avalanche size distributions for the subcritical $(\lambda<1)$, critical $(\lambda=1)$, and supercritical $(\lambda>1)$ regime (fig.\ref{fig1} b-d). 

To investigate the hypothesized impact of the network state on the statistics of neuronal spiking, we simulated the network activity for different values of $\lambda$ and quantified the single-neuron spiking statistics using the CV of the inter-spike-interval distributions (fig.\ref{fig1} e). The CV is defined as the ratio of the standard deviation and the mean of the inter-spike-interval distribution for a given neuron. The irregular spiking is basically characterized by $CV>1$, whereas $CV=1$ is considered as Poisson spiking. It turned out that at the presence of a constant slow drive (see \nameref{methods}) the CV values distributed around a mean greater than one, thus indicating irregular spiking. 
In contrast, small deviations of the network state towards either the subcritical or the supercritical regime resulted in CV values distributed around a mean of 1 or less. In summary, when tuning the network from the subcritical to the supercritical state, the mean CV peaked near $\lambda=1$ (fig.\ref{fig1}f); with increasing network size, the peak moved closer towards $\lambda=1$ and becomes narrower as well. This observation suggests that, at the large-size limit, the irregular spiking is an emergent property of recurrent networks operating at the critical point ($\lambda=1$). In other words, the scale-free fluctuations in network activity at criticality translate into irregular single-neuron spiking. 

\begin{figure*}
\centering
\includegraphics[scale=0.42]{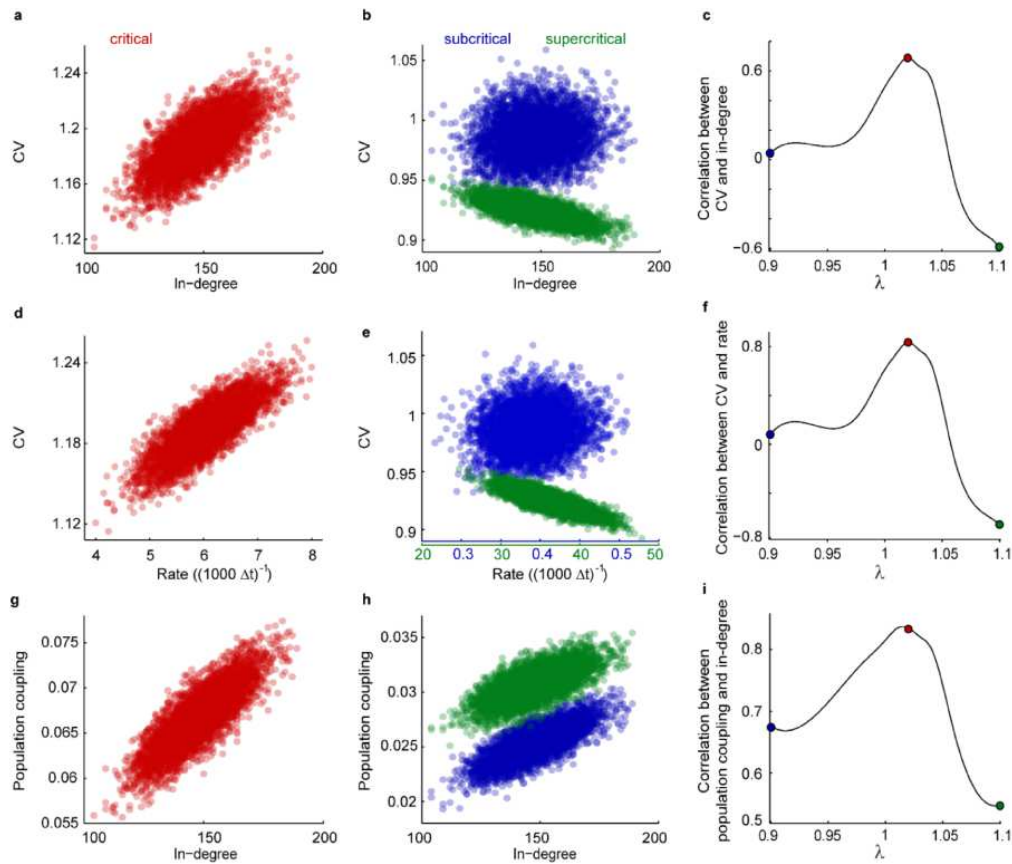}
\caption[{\protect \footnotesize Solely for a recurrent network operating near criticality does the irregularity of spiking increase with a neuron's in-degree:}]{\protect \footnotesize {{\bf Solely for a recurrent network operating near criticality does the irregularity of spiking increase with a neuron's in-degree:} {\bf (a, b)} The inter-spike-interval CVs from simulated spike trains versus the neuron's in degree for a model network in three different states: (a) critical $(\lambda=1.02)$, (b) subcritical $(\lambda=0.9)$ and supercritical $(\lambda=1.1)$. Network parameters were $N=5000$, $3\%$ connectivity, and $\eta=1/(5N)$ applied to all neurons. {\bf (c)} Correlation between CV and in-degree as a function of $\lambda$. Other network parameters as in (a, b). The curves were smoothed using Matlab spline. The colored dots correspond to the distributions in (a, b). {\bf (d, e)} $CV$ versus rate for three network states. Network parameters as in (a, b). (f) Correlation between CV and rate as a function of $\lambda$. Other network parameters as in (a, b). {\bf (g, h)} Population coupling versus in-degree for three network states. Network parameters as in (a, b). (i) Correlation between population coupling and in-degree as a function of $\lambda$. Other network parameters as in (a, b).} }
\label{fig2}
\end{figure*}
The above results could be confounded by the finite network size. Importantly, the width of the mean $CV$ as a function of $\lambda$ decreased with increasing network size and the peak location moved closer to $\lambda=1$ (fig.\ref{fig1}f). To further test whether the deviation of the peak from $\lambda=1$ is indeed due to the finite size effect, we looked at two major characteristics of criticality, namely maximum correlations and mean response. In order to see how the average correlation among neurons behaves in terms of the control parameter $\lambda$, we computed a commonly-used measures of coordinated network activity that is known to be maximized at criticality. One such quantity is the {\it average population coupling}, which represents a measure of the overall level of correlated fluctuations within the network \cite{okun2015diverse} (see \nameref{methods}). 

Comparing the CV and the average population coupling reveals that they both peak at the same point (close to $\lambda=1$), also similar to the change in mean response to an increase in external input. 
This demonstrates that the peak for CV coincides with the critical point, which is characterized by maximum average population coupling and change in mean response. Therefore, maximum CV could be regarded as a hallmark of critical dynamics for recurrent neural networks.

The observed onset of irregular spiking near criticality (fig.\ref{fig1}f) has an intuitive explanation in the extreme limits of connection strength. In the limit of weak connections $(\lambda<1)$, spiking is largely driven by the Poisson external input alone, thus resulting in Poisson like spiking. On the other hand, neurons become mostly active leading to more regular spiking in the limit of very strong connections $(\lambda>1)$. However, it is at criticality $(\lambda \approx 1)$ that the scale-free fluctuations in network activity translate to single-neuron irregular spiking $(\langle CV \rangle > 1)$. Furthermore, although there can be alternative mechanisms for irregular spiking, such as synchronous inputs (see discussion), we show that the impact of the network state on the statistics of single-neuron spiking is largely robust with respect to structured external inputs; the irregularity of spiking is always maximized near network criticality even at the presence of other sources of irregular spiking (fig.\ref{figS1},fig.\ref{figS2},fig.\ref{figS3}). 
\begin{figure*}
\includegraphics[scale=0.35]{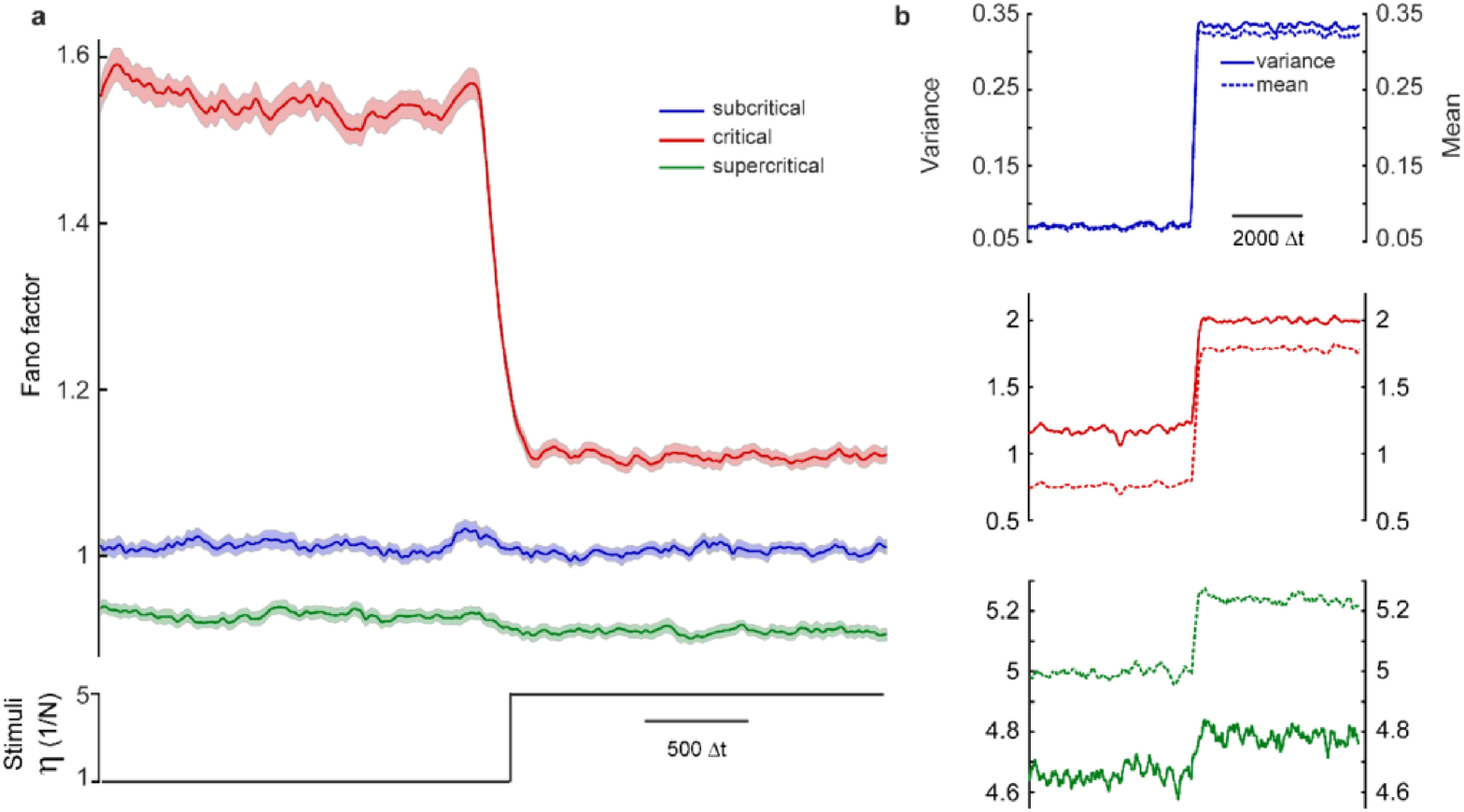}
\caption[{\protect \footnotesize The network state determines the change in response variability after stimulus onset.}]{\protect \footnotesize {{\bf The network state determines the change in response variability after stimulus onset. } {\bf (a)} Average Fano factor (solid lines) from repeated step increases in the external input for three different network states: subcritical ($\lambda=0.95$, blue), critical ($\lambda=1.02$, red), and supercritical ($\lambda=1.07$, green). Spike trains were simulated for a network with $N=5000$ neurons and $3\%$ connectivity. External input was applied to $10\%$ of the neurons and external input strength switched from $\eta=1/N$ to $\eta=5/N$ as indicated. The Fano factor is the ratio of the variance and mean in the number of spikes within a given time window and repeated trials. The sliding window was $200$ time steps and the sliding increment was $20$ time steps. We calculated the average Fano factor from 60 randomly sampled model neurons and $2000$ repeated trials. {\bf (b)} The mean and variance of the spike rate when the average external input strength switched from $1/N$ (first half) to $\eta=5/N$ (second half) for three different network states. All network parameters as in (a).} }
\label{fig3}
\end{figure*}
The observed impact of the network state on the statistics of single-neuron spiking ( fig.\ref{fig1}e,f) raised the question to what extent this network-to-neuron impact is regulated by a neuron's in-degree, i.e., the number of input connections to a neuron. To address this question, we took advantage of the distribution of in-degrees provided by a model network with sparse and random connectivity. 

We found that near criticality $(\lambda \approx 1)$, a neuron's CV tended to increase with increasing in-degree (fig.\ref{fig2}a). Importantly, however, this correlation between a neuron's activity statistics (CV) and its connectivity (in-degree) changed drastically when tuning the network state away from criticality. In the subcritical regime only weak correlation was found and in the supercritical regime the relation reversed, i.e., a neuron's CV tended to decrease with increasing in-degree (fig.\ref{fig2}b). In general, when tuning the network state through the critical regime, the relation between a neuron's CV and its in-degree transforms from small correlation in the subcritical regime $(\lambda < 1)$, to maximized correlation at the critical point $(\lambda \approx 1)$, and to anti-correlation in the supercritical regime $(\lambda >1)$,  (fig.\ref{fig2}c). 

The relationship that solely near criticality is the irregular spiking maximally reflective of the neuron's in-degree, suggests a novel measure to test the criticality hypothesis in electrophysiological experiments. It is of practical importance that in a network dominated by excitatory neurons, a neuron's firing rate scales with its in-degree and that this relation is independent of the network state. Thus, spike train recordings from a population of neurons can be informative about the network state: a maximum correlation between a neuron's CV and its firing rate is indicative of a critical network state, whereas weaker correlation or anti-correlation is indicative of the subcritical or supercritical network state, respectively (fig.\ref{fig2}d-f). 

In comparison, the relation between a neuron's "population coupling" (see \nameref{methods}) and its in-degree is less decisive about the network state, as population coupling increases with a neuron's in degree for all three network states (fig.\ref{fig2}g, h). However, this relation is also dominant for the critical network state (fig.\ref{fig2}i). Consistent with these model results, recent experimental data from in vivo recordings showed increased population coupling with increasing 
synaptic inputs \cite{okun2015diverse}. However, unless the network state is manipulated and a maximum in the correlation between the population coupling and the in-degree is determined, that data set is not informative about the network state. \\

In addition to the irregular spiking evaluated above, a decline of response variability at stimulus onset is a prominent cortical phenomenon \cite{churchland2010stimulus}. The important question that how an external input suppresses the variability of ongoing activity has remained intriguing \cite{sussillo2009generating,rajan2010stimulus}. To address this question in the context of the network state, we simulated spiking activity of the model network near criticality $(\lambda \approx 1)$ in response to repeated increases in the external input. We quantified the across-trial firing rate variability in terms of the {\it Fano factor}, which is a measure of the across-trial variability in the number of spikes in relation to the trial-averaged mean number of spikes during a given window of time (see \nameref{methods}). The Fano factor has a value of approximately 1 for repeated Poisson spike trains. Recordings from cortical neurons show a Fano factor above 1 for ongoing activity and a significant drop during external stimulation \cite{churchland2010stimulus}. 
Our simulations of a model network in the critical state reproduced this drop in Fano factor from a high value during ongoing activity to a lower value (still above 1) during external stimulation (fig.\ref{fig3}a). In contrast, in the subcritical and the supercritical network state, the Fano factor remained unchanged when switching from ongoing to evoked activity. This result demonstrates that the experimentally observed drop in Fano factor could naturally emerge as 
a characteristic feature of the critical network state. The result is qualitatively robust with respect to the details of the external inputs (fig.\ref{figS4}). 

Since the Fano factor is defined as a ratio of the variance divided by the mean, the declining Fano factor at criticality could result trivially from an increased mean spike count accompanied by a weak dependence of the variance on the network state during external stimulation. Evaluating the changes in the spike count and the variance separately ruled out this possibility (fig.\ref{fig3}b). The "mean spike count" and the variance increased with external stimulation for all three network states. However, only near criticality, when the network is maximally sensitive to external stimuli, did the mean spike count increase more drastically, thus yielding a reduced Fano factor. In conclusion, our model simulation suggests that the experimentally observed drop in Fano factor after stimulus onset \cite{ churchland2010stimulus} can be the result of the cortical circuit operating at criticality. 

\protect \section{Discussion}  \label{discussion}
A wealth of evidence indicates that neural networks of cerebral cortex operate at criticality \cite{ schuster2014criticality,arviv2015near,shew2015adaptation}. Here, we showed that (i) irregular spiking \cite{softky1993highly,shadlen1998variable,bellay2015irregular}, (ii) its relation to the neuron's in-degree, and (iii) the decline of response variability at stimulus onset \cite{ churchland2010stimulus}, are all emergent properties of a recurrent network operating at criticality. A number of separate dynamical, biophysical, and structural mechanisms have been proposed to generate the observed irregular spiking \cite{sussillo2009generating,van1996chaos,stevens1998input,kumar2008conditions,litwin2012slow, ostojic2014two}. The significance of our work resides in part in establishing criticality as one unifying principle for both the collective behavior of cortical circuits and the statistics of single-neuron spiking. Indeed, experimental evidence for the predicted coexistence of irregular spiking and criticality has recently been provided in recordings of ongoing cortical activity in vivo \cite{bellay2015irregular}. While the coexistence of irregular spiking and power-law avalanche size distributions has been demonstrated in more complex model networks\cite{chen1995self,eurich2002finite,benayoun2010avalanches,millman2010self,stepp2015synaptic}, our work extends qualitatively beyond these important earlier studies in three fundamental dimensions. First, the choice of a network of excitatory probabilistic integrate-and-fire model neurons allows the precise analytic evaluation of the network state via a single control parameter $\lambda$, i.e., the maximum eigenvalue of the transfer probability matrix. This approach avoids the need to rely on the cumbersome and less precise avalanche analysis to evaluate the network state. Second, our discovery that the relation between the irregularity of spiking and the neuron's in-degree/firing rate is maximized at criticality provides an important new measure of criticality. Furthermore, this relation establishes a valuable conceptual link between criticality and the important field of network theory, where a node's in-degree is a basic system parameter. Third, as the observed decline in response variability is regarded as an essential mechanism to enhance response fidelity to stimuli \cite{ churchland2010stimulus}, our discovery of its relation to network criticality offers a starting point toward unraveling the possible roles of critical dynamics in neural coding. In conclusion, it will be interesting to see to what extent the presented findings will generate a paradigm shift in the study of criticality of neural systems: our results build a much-needed bridge between critical dynamics and neural coding, and provide novel and robust measures to test the criticality hypothesis itself.\\

\section{Methods} \label{methods}
{\small{ 
We simulated a model network consisting of excitatory binary probabilistic model neurons with sparse connectivity and external inputs. Network size ranged from 5000 to 20000 model neurons. The strength of the connection from neuron $j$ to neuron $i$ is quantified in terms of the transition probability $P_{ij}$, which is the probability that a spike in neuron $j$ causes a spike in neuron $i$ in the next simulation time step. For a network of $N$ neurons and an average connectivity $K$, each neuron is connected to $N-1$ other neurons with probability $K/N$. For each (on average) $K(N-1)$ connections among neurons a $P_{ij}$ is assigned by drawing a random number from uniform distribution in the interval $[0 \, \, \frac{2}{K}]$. With sufficiently large  this yields a network with a normally distributed connectivity with average $K$ and a transition matrix $P_{ij}$ with maximum (absolute value) eigenvalue of 1. In order to deviate the network from the critical point we can simply multiply $P_{ij}$  by a factor smaller(greater) than one. The binary state $X_i(t)$ of neuron $i$ denotes whether the model neuron spikes $(X_i(t)=1$ or does not spike $(X_i(t)=0$ at time $t$. At each time step, the state of all neurons were updated synchronously according to the following update rule: 
\begin{equation}
X_i(t+1) = \Theta \bigg[ \Big(1-\eta_i(t)\Big) \sum_j P_{ij} X_j(t) + \eta_i(t) - \xi_i(t) \bigg]
\end{equation}

where $\eta_i(t)$ of neuron $i$ quantifies the probability of that neuron spiking only due to external input, $\xi(t)$ is a random number in $[0 1]$ drawn from a uniform distribution, and $\Theta$ is the step function. In addition to the update rule, a refractory period of two time-steps was implemented. The external input $\eta_i(t)$ was chosen to be smaller than the transition probability $P_{ij}$, which itself was small for large networks, $P_{ij}\sim 1/N$. Because of the weak external inputs, we employed the approximation $1-\eta \approx 1$ in the above update rule. The maximum eigenvalue $\lambda$ of the transition probability matrix $P_{ij}$  describes the network state at the thermodynamic limit: $\lambda<1$ denotes subcritical regime, $\lambda \approx 1$  denotes the near critical regime and $\lambda>1$  denotes the supercritical regime. However, for finite-sized networks we evaluate the exact critical point based on the peak of average population coupling (see below). The $P_{ij}$ values were drawn from a uniform distribution and then scaled by constant to reach the desired maximum eigenvalue $\lambda$. The constant external input $\eta_i(t)$ was either constant. 

The external input $\eta_i(t)$ was either constant or was modeled as a binary Poisson process followed by smoothing with a Gaussian filter with a width of 20 time steps and multiplied by an amplitude factor $\eta_0$ between 0 and 1. The synchronous input (fig.\ref{figS1}) was simulated with replicating a single binary Poisson process smoothed with a Gaussian filter with a width of 100 time-steps. In order to apply some variability among the stimuli received by different neurons each smoothed Poisson process was multiplied by a factor of $\eta_0 + 0.2\epsilon$ where $\epsilon$ was drawn from a normal distribution. 

We analyzed the simulated spike trains with respect to five complementary statistical measures: neuronal avalanches, coefficient of variation, population coupling, change in mean response, and Fano factor. Following commonly-used procedures \cite{beggs2003neuronal}, a neuronal avalanche was defined as an episode of continuous (each time step) network spiking, framed by time steps of no spikes. Avalanche size was taken as the number of spiking neurons. The single-neuron spike train variability was quantified using the coefficient of variation $CV \equiv \sigma_{isi}/\mu_{isi}$, defined as the ratio of the standard deviation $\sigma_{isi}$ and the mean $\mu_{isi}$ of the inter-spike-interval $(ISI)$ distribution for a given neuron. We managed simulation times to be sufficiently long to ensure stable $CV$ values. The coordination within the network was quantified using the population coupling , which is defined as the zero-lag cross-correlation between the spike train $X_i(t)$ of neuron $i$ and the remaining network activity $N_i(t)=\sum_{j\ne i}X_j(t)$ from all other spike trains, i.e., $ c_i=\frac{ \Big\langle \big(X_i(t)-\langle X_i(t)\rangle \big) \big(N_i(t)-\langle N_i(t) \rangle \big) \Big\rangle }{\sigma_X \sigma_N} $, where the angled brackets indicate a time average \cite{okun2015diverse}. Averaging the population coupling  over many neurons within a large network yields the average population coupling , which represents a measure of the overall level of coordination within the network. The change in mean response was computed as the difference in the mean spike counts for external inputs of $\eta=1/(5N)$  and $\eta=2/N$. When quantifying the spike train variability in the context of repeated experimental situations, it is convenient to use the Fano factor, which, for a given time window and repeated trials, is defined as the ratio of the variance and mean in the number of spikes. We chose a window of $200$ time-steps (sliding in increments of $20$ time steps) and repeated trials $2000$ times. We calculated the average Fano factor from $60$ randomly sampled model neurons. 
}\\

{\bf Acknowledgment} 
We thank Anders Carlsson, John Clark, and Woodrow Shew for comments on previous versions of the manuscript. This research was supported by a Whitehall Foundation grant $\#20121221$ (R.W.) and a NSF CRCNS grant $\#1308159$ (R.W.).
		

\bibliographystyle{ieeetr}
\bibliography{./References-arxiv}

\nocite{*}

%

\protect \widetext
\protect \clearpage
\begin{center}
\protect \textbf{\large Supplementary Figures}
\end{center}
\protect \setcounter{figure}{0}
\makeatletter
\renewcommand{\theequation}{S\arabic{equation}}
\renewcommand{\thefigure}{S\@arabic\c@figure}
\renewcommand{\thesuppfigure}{S\@arabic\c@suppfigure}
\renewcommand{\thefigure}{S\arabic{figure}}%

\begin{figure*}[h]
\centering
\includegraphics[scale=0.6]{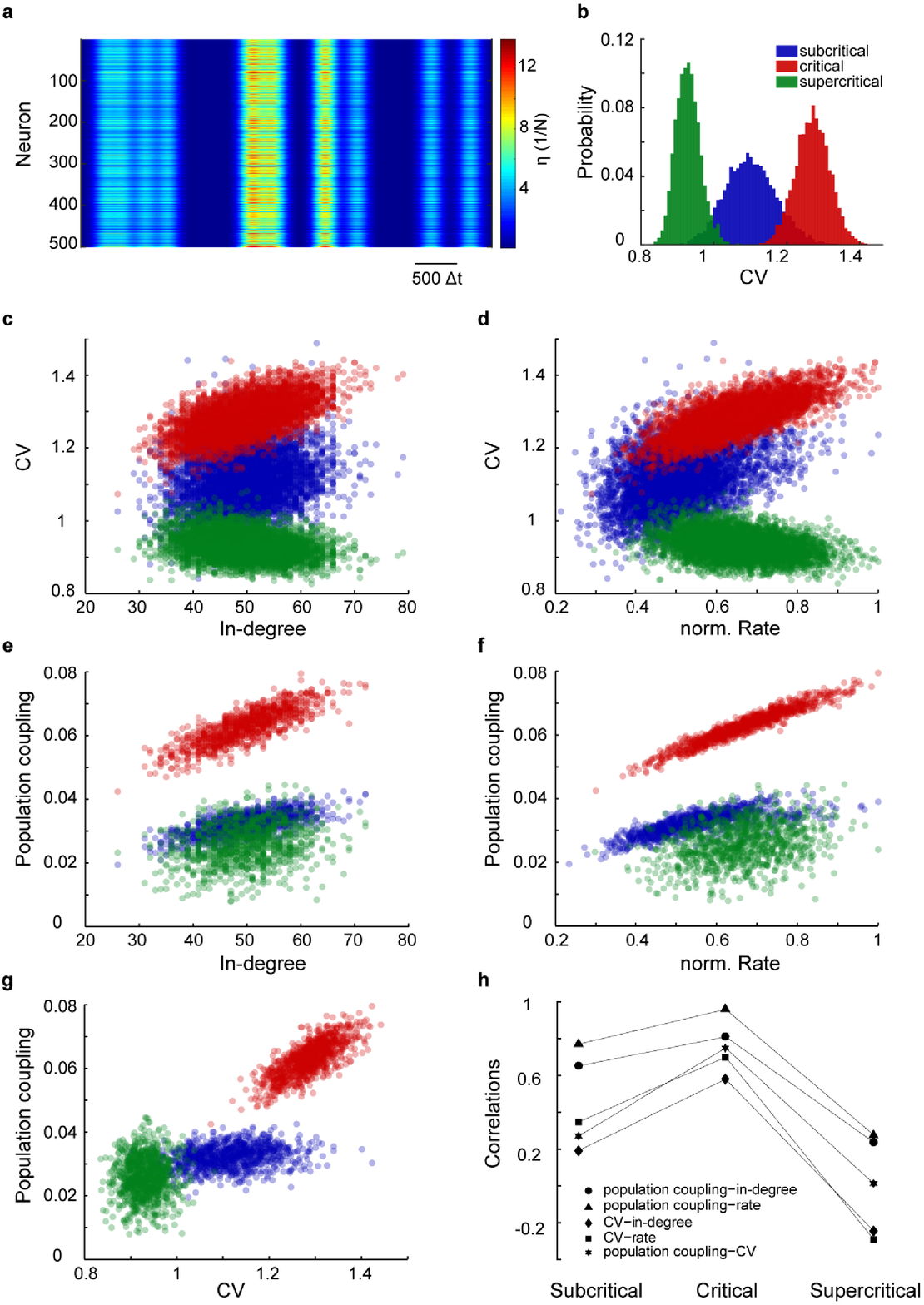}
\caption[{\protect \footnotesize Irregular spiking at criticality in a recurrent network with synchronous external inputs.}]{\protect \footnotesize {{\bf Irregular spiking at criticality in a recurrent network with synchronous external inputs. } 
{\bf (a)} The temporal structure and strength of the external input $\eta(t)$ to $10\%$ of the neurons in a recurrent model network of $5000$ neurons and $1\%$ connectivity. The external input $\eta(t)$ was generated from Poisson pulses of rate $10/N$, smoothed by a Gaussian filter of width $100$ time-steps and amplitude of $0.2(1+\epsilon)$, where $\epsilon$ is drawn from a normal distribution (see \nameref{methods}). This synchronous input was added to a background constant external input of $1/(10N)$. {\bf (b)} Inter-spike-interval CV distributions of simulated spike trains for the subcritical ($\lambda = 0.95$, blue), critical ($\lambda = 1.09$, red), and supercritical ($\lambda = 1.09$, green) network state. At the critical regime the spike trains show highest irregularity, which is indicated by the peak of the CV distribution located near 1.3.  {\bf (c, d)} The inter-spike-interval CVs from simulated spike trains versus the neuron's in degree (c) and its normalized rate (d) for the three network states. {\bf (e, f)} The population coupling from simulated spike trains versus the neuron's in degree (e) and its normalized rate (f) for the three network states. {\bf (g)} The population coupling vs a neuron's CV for the three network states. {\bf (h)} The Spearman correlation coefficients between CV and in-degree (rate), population coupling and in-degree (rate), population coupling and CV are all maximized near criticality. } }
\label{figS1}
\end{figure*}

\newpage
\begin{figure*}[h]
\centering
\includegraphics[scale=0.65]{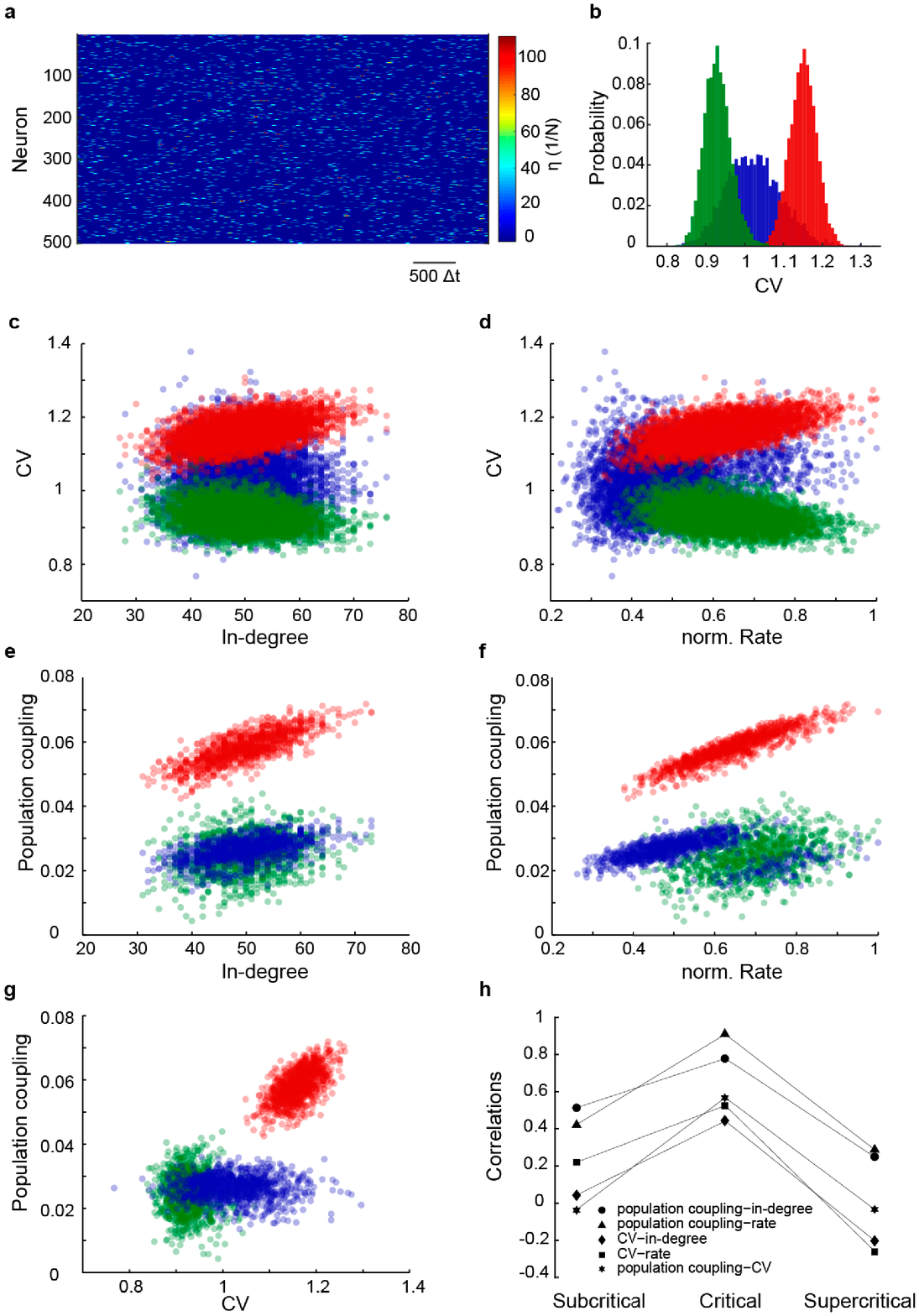}
\caption[{\protect \footnotesize Irregular spiking at criticality in a recurrent network with asynchronous external inputs}]{\protect \footnotesize { {\bf Irregular spiking at criticality in a recurrent network with asynchronous external inputs.} 
{\bf (a)} The temporal structure and strength of the external input $\epsilon$ to $10\%$ of the neurons in a recurrent model network of 5000 neurons and $1\%$ connectivity. The external input $\epsilon$ was generated by independent Poisson pulses of rate $5/N$, smoothed by a Gaussian filter of width 20 time-steps and amplitude $\eta_0 = 0.5$ (see \nameref{methods}). {\bf (b)} Inter-spike-interval CV distributions of simulated spike trains for the subcritical ($\lambda = 0.95$, blue), critical ($\lambda = 1.02$, red), and supercritical ($\lambda = 1.09$, green) network state. {\bf (c, d)} The inter-spike-interval CVs from simulated spike trains versus the neuron's in degree (c) and its normalized rate (d) for the three network states. {\bf (e, f)} The population coupling from simulated spike trains versus the neuron's in degree (e) and its normalized rate (f) for the three network states. {\bf (g)} The population coupling vs a neuron's CV for the three network states. {\bf (h)} The Spearman correlation coefficients between CV and in-degree (rate), population coupling and in-degree (rate), population coupling and CV are all maximized near criticality.  } }
\label{figS2}
\end{figure*}


\begin{figure*}[h]
\centering
\includegraphics[scale=0.4]{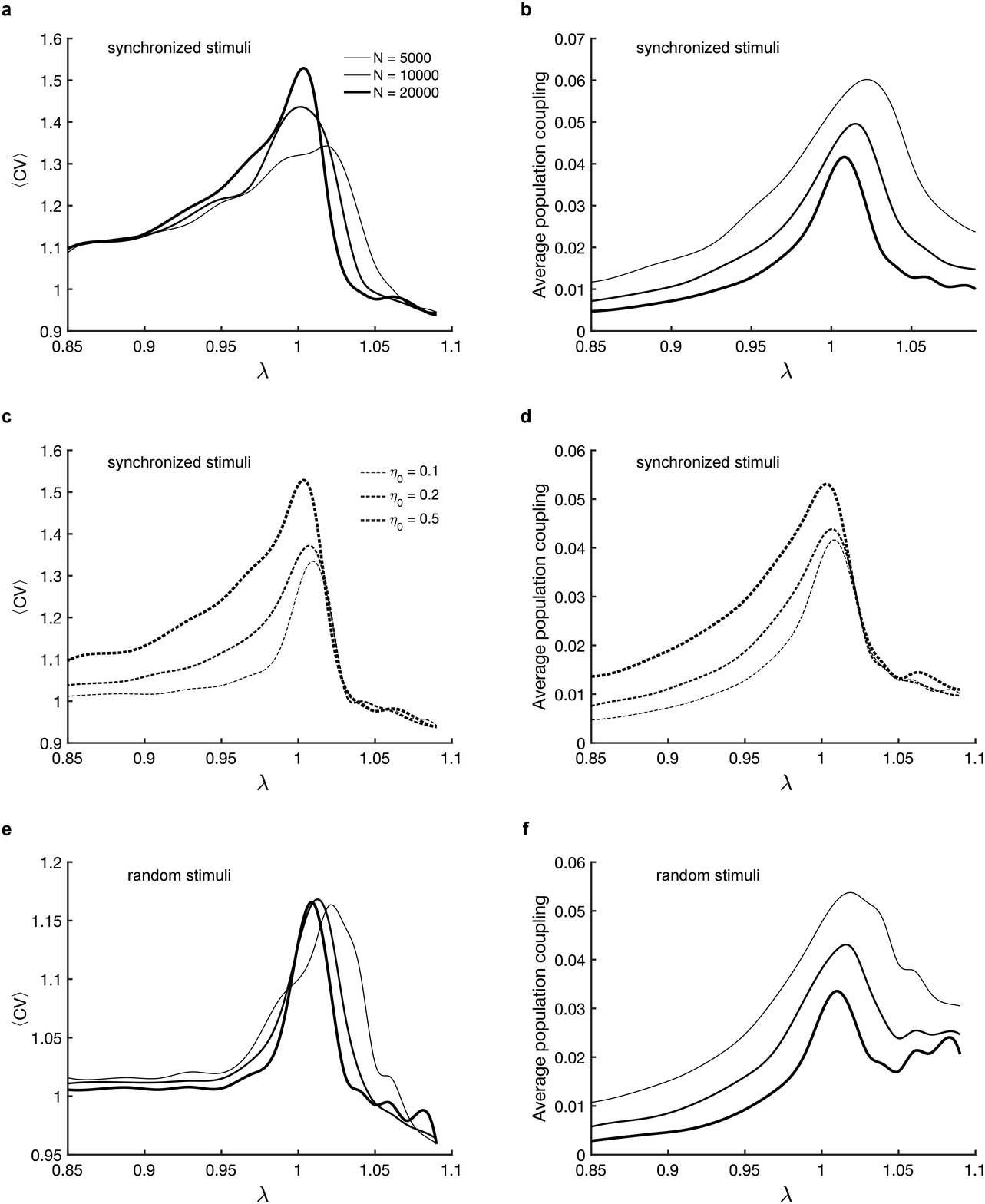}
\protect \caption[{\protect \footnotesize The average CV and average population coupling are maximized near network criticality for external inputs of different spatiotemporal structure.}]{\protect \footnotesize {\bf The average CV and average population coupling are maximized near network criticality for external inputs of different spatiotemporal structure. }
{\bf (a, b)} Average CV (a) and average population coupling (b) vs the control parameter ?? for synchronous external inputs (see Fig. S1a, but with different stimulation amplitudes (see fig.\ref{figS1}a, but with stimulation amplitude $\eta_0 = 0.1 , 0.2 , 0.5$; see \nameref{methods}) for three different network sizes. {\bf (c, d)} Average CV (c) and average population coupling (d) vs the control parameter $\lambda$ for synchronous external inputs (see fig.\ref{figS1}a) for a network size of $N=5000$ and for three different stimulus amplitudes. {\bf (e, f)} Average CV (e) and average population coupling (f) vs the control parameter $\lambda$ for asynchronous external inputs (see fig.\ref{figS2}a) for three different network sizes (see legend in (a)). }
\label{figS3}
\end{figure*}


\begin{figure*}[]
\centering
\includegraphics[scale=0.8]{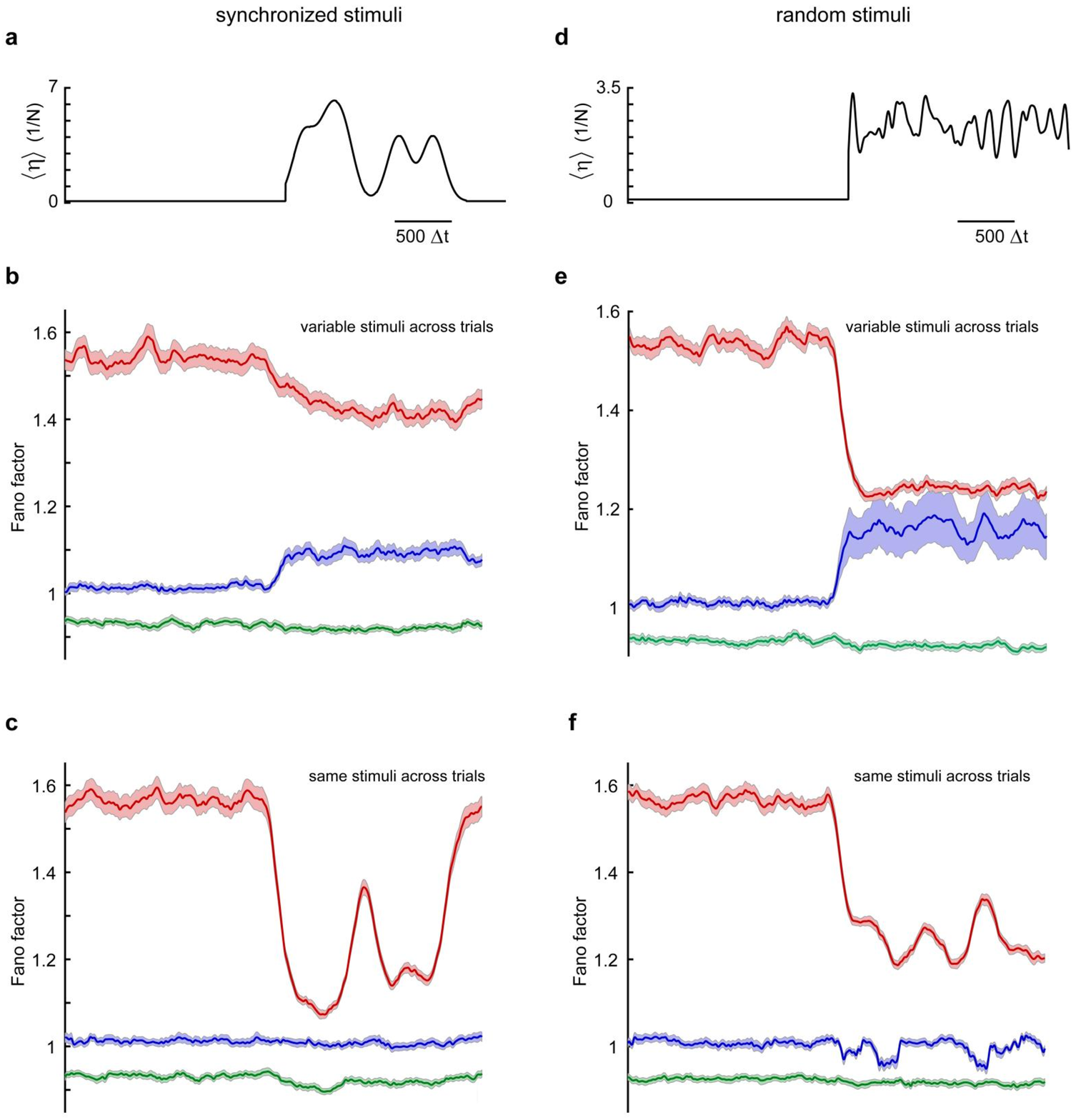}
\caption[{\protect \footnotesize The change in response variability for external inputs of different temporal structure.}]{\protect \footnotesize {\bf The change in response variability for external inputs of different temporal structure.}
{\bf (a)} Average external input for the case of synchronous external inputs applied to $10\%$ of the neurons as described in fig.\ref{figS1}a. {\bf (b)} The average Fano factor computed for 60 neurons over 2000 trials of different stimuli, but of the same type. In the presence of temporally structured stimuli, the Fano factor can increase at sub-criticality, due to the across-trial variability in the stimuli. {\bf (c)} The average Fano factor computed for the same neurons over 2000 trials of the exactly the same stimulus (as is shown in (a). {\bf (d-f)} Same as (a-b) but for asynchronous external inputs as described in fig.\ref{figS2}a. }
\label{figS4}
\end{figure*}


\end {document}